\documentclass[12pt,]{article}
\usepackage{amsfonts}
\usepackage{amsmath,amssymb,bm}
\usepackage{graphicx}

\input{tcilatex}

\begin{document}

\title{Incompressible fluid inside an astrophysical black hole?}
\author{F. Canfora \\
{\small \textit{Centro de Estudios Cientificos (CECS), Casilla 1469
Valdivia, Chile.}}\\
{\small \textit{Istituto Nazionale di Fisica Nucleare, Sezione di Napoli, GC
Salerno, Italy.}}\\
{\small e-mail: \textit{canfora@cecs.cl}}}
\date{}
\maketitle

\begin{abstract}
It is argued that under natural hypothesis the Fermions inside a black hole
formed after the collapse of a neutron star could form a non compressible
fluid (well before reaching the Planck scale) leading to some features of
integer Quantum Hall Effect. The relations with black hole entropy are
analyzed. Insights coming from Quantum Hall Effect are used to analyze the
coupling with Einstein equations. Connections with some cosmological
scenarios and with higher dimensional Quantum Hall Effect are shortly
pointed out.
\end{abstract}


Preprint: CECS-PHY-07/17 \newline

\section{Introduction}

Black-hole thermodynamics and Hawking radiation (\cite{Ba73} \cite{Be73} 
\cite{Ha74}) are some of the few sound results in which general relativity
and Quantum Field Theory (henceforth QFT) fit together. On the other hand,
there are some important open problems in this field. Mainly, it is still
not understood the final stage of black hole evaporation and how to solve
the information loss paradox. Black hole entropy is proportional to the area
of the horizon and many different ways to deduce such an area law have been
proposed (for a review see \cite{Carl07} and references therein). Many of
such models assume that the effective degrees of freedom of the black hole
live on the boundary of the black hole itself and can be described by a
conformal theory. Indeed, the success of such proposals has partially
inspired the \textit{Holographic Principle} (after the pioneering ideas of
Bekenstein \cite{Be81}, 't Hooft \cite{tH93} and Susskind \cite{Su95}) which
is believed to play a fundamental role in the yet to be discovered final
theory of gravity. However, the questions of \textit{why such effective
degrees of freedom should live on the boundary} of the black hole and of 
\textit{how} \textit{the bulk degrees of freedom get frozen} are still
opened.

Here it is analyzed the case of a black hole formed due to the collapse of a
typical neutron star. It is argued that at an energy scale of order $%
10^{-21}\sim 10^{-18}$ of the Planck scale (after the horizon is formed)
many features of Quantum Hall Effects (henceforth QHE) come into play. The
classical gravitational force is likely to dominate the other processes of
the standard model and the spectrum of the Fermions living inside the black
hole turns out to be discrete. Due to the gap, the Fermions gas inside the
black hole cannot be compressed anymore. It is not a scope of the present
paper neither to write down an effective action for the effective degrees of
freedom of a black hole nor to argue about the quantum degrees of freedom of
gravity. The idea is to explain \textit{why}, in a concrete situation, many
of the available "conformal" descriptions of the effective degrees of a
black hole \textit{should work}.

Based on the AdS/CFT correspondence \cite{Ma97} and on the geometry of the
3-dimensional BTZ black hole \cite{Banados:1992gq} \cite{Banados:1992wn}, an
analogy between Quantum Hall effect and gravity in three dimensions has been
pointed out in \cite{Myu98}. It appears that the Quantum Hall bulk degrees
of freedom as well as the edge excitations are suitable to describe the
dynamical features of the BTZ black hole. The perspective in \cite{Myu98} is
completely different from the present scheme in which the starting point is
the collapse of a neutron star in 4-dimensional general relativity. It is
therefore interesting that many connections between so different approaches
arise anyway.

The structure of the paper is the following: in Section 2, the assumptions
of the present paper are explained and the standard order of magnitude
inside a\ neutron star are resumed. In Section 3 the arising of features
typical of Quantum Hall Effects is described. In Section 4 the relations
with black hole entropy are analyzed and a simple bound on the entropy is
derived. In Section 5 the Einstein equations in the presence of a "Quantum
Hall" source are solved and the connection with higher dimensional Quantum
Hall formalism is pointed out. In Section 6 the relations of the present
proposal with some interesting cosmological scenarios are outlined together
with the possible weakness of the approach. In Section 7 some conclusions
are drawn.

\section{The standard approximations}

The first basic assumption of the standard model of a neutron star is that
at the energy scale of the standard model of particles physics (up to $TeV$)
the collapsing neutron star can be described very well by QFT coupled to
classical general relativity.

The second basic assumption is that the quantum dynamics of the neutrons (or
of the quarks) living inside the neutron star is much faster than the
dynamics of the gravitational field. This implies that one can compute the
equation(s) of state of the Fermions as usual and then use such equation(s)
to solve the Einstein equations in which the source is described by the
equation of state itself. The success of this theory initiated by Landau,
Chandrasekhar, Tolman, Oppenheimer, Volkoff, Snyder (and many others) tells
that such adiabatic approximation is excellent.

The two basic assumptions which will be needed in the following are that the
above approximations also hold up to an energy scale of $10^{-21}-10^{-15}$
of the Planck scale: it will be assumed that the standard model and general
relativity are the correct theories at these scales. It will be also assumed
that at this scale of energy the Fermions living inside the collapsed
neutron star have a dynamic much faster than the dynamics of the
gravitational field so that one can compute the equation of state of the
Fermions and than use the result to solve the Einstein equations coupled
with the Fermions themselves.

\subsection{Orders of magnitude inside a neutron star}

Let us remind the typical order of magnitudes of a Neutron Star (henceforth
NS). The typical mass $M_{NS}$, radius $R_{NS}$, density $\rho _{NS}$, the
baryon number $N_{NS}$ and the Schwarzschild\ radius $r_{G}(M_{NS})$ of a NS
are%
\begin{eqnarray*}
M_{NS} &\approx &10^{33}g,\quad R_{NS}\approx 10^{6}cm,\quad \rho
_{NS}\approx 10^{15}g/cm^{3},\text{ } \\
N_{NS} &\approx &10^{54},\quad r_{G}(M_{NS})\approx 10^{4}cm.
\end{eqnarray*}%
One can compute the strength of the gravitational interaction on a Fermions
living inside a neutron star (which to a very good approximation can be
considered as a sphere of constant density) of these characteristic, the
result is%
\begin{equation*}
\hslash \omega _{G}=\hslash \sqrt{G\rho _{NS}}\approx 10^{-38}E_{Planck}.
\end{equation*}%
Therefore, the gravitational interaction is negligible when compared to the
strength of the interactions of the standard model (which are of the order
of $10^{-21}\sim 10^{-18}$ of the Planck scale). A neutron star with a mass
of the above order of magnitude is unstable against the gravitational
collapse to a black hole. During the collapse the baryon number is conserved
so that the black hole which is eventually formed should have the same
parameters $M_{NS}$ and $N_{NS}$ of the parent neutron star (let us forget
for a moment Hawking radiation which is, in any case, negligible for black
holes of the mass of a neutron star; the issues connected with Hawking
radiation will be shortly discussed later on). On the other hand, at a first
glance, $R_{NS}$ should decrease without bound at least up to the Planck
length since, apparently, there is no process which can prevent such a
decreasing after the black hole is formed since "gravity dominates Pauli
pressure". In fact, if the radius decreases, at a certain point the strength
of the gravitational interaction on a Fermions inside a Black hole will be
comparable and stronger than the other interactions. When the radius
decreased up to the following value of the density%
\begin{equation}
\hslash \omega _{G}^{\ast }=\hslash \sqrt{G\rho _{NS}^{\ast }}\approx \left(
10^{-21}-10^{-18}\right) E_{Planck}  \label{0QHC}
\end{equation}%
the other interactions among the Fermions inside the black hole should be
treated as perturbations of the gravitational interaction.

\section{Quantum Hall Effect and black holes}

The principal insight comes from the physics of Integer QHE but possible
effects related to the interactions of the Fermions which could give rise to
phenomenology of the Fractional QHE should not be excluded \textit{a priori}%
\cite{Lau99}. In the presence of a strong confining potential a gap opens up
in the spectrum. Therefore, when the number of Fermions is such that an
integer number of levels is full the gas becomes incompressible because of
the gap so that its equation of state is simply $\rho =const$. If the
assumptions made above are correct this also should happen inside a black
hole formed after the collapse of a neutron star. The gravitational field
inside a neutron star can be well approximated by the Newtonian harmonic
oscillator\footnote{%
Indeed, the first computations in the theory of the gravitational
equilibrium of a neutron star, made in these approximations, were quite
successful. The reason is that the neutrons perceive around them a density
which is almost uniform. Moreover, the probability for a neutron to escape
from the star is negligible already at the level of Newtonian gravity and,
as it is well known, "general relativity is more attractive" than Newtonian
gravity.} of frequency $\omega _{G}=\sqrt{G\rho _{NS}}$. As it will be
discussed in a moment, the corrections due to general relativity enhance the
arising of "Quantum Hall phenomenology" so that the essential physics can be
understood using the Newtonian expression of the gravitational field. The
gap satisfies%
\begin{equation}
\hslash \sqrt{G\rho _{NS}^{\ast }}=E_{gap}>E_{SM}  \label{1QHC}
\end{equation}%
where $E_{SM}$ is the typical energy scale of a process of the standard
model.

That this should happens well before to reach the Planck scale can be argued
as follows. Any cross section $\sigma _{SM}$ computed in the standard model%
\footnote{%
In the case of a neutron star, at the densities at which gravity begins to
dominate, it is possible that all the neutrons could be transformed into
quarks. In this case, besides gravity, the dominant interaction would be the
strong interaction which is asymptotically free.} decreases with energy
(because of the unitarity of the gauge interactions appearing in the
standard model):%
\begin{equation*}
\sigma _{SM}\sim g(s)s^{-\gamma },\quad \gamma >0
\end{equation*}%
where $s$ is the typical energy (which in this case is proportional to a
negative power of the radius of the collapsed neutron star) and $g(s)$ is
the coupling constant at the energy scale of interest\footnote{%
In the case of the gauge interactions of the standard model the "worst" case
could be one in which the coupling constant increases logaritmically with
the energy scale. However, this behavior does not change the main
qualitative conclusion that at an energy scale well below the Planck scale
inside a neutron star gravity begins to dominate. Moreover, the results in 
\cite{RW05} indicate that gravitational corrections lower the scale of
asymptotic freedom of the gauge interactions.}. On the other hand, the
strength of the gravitational oscillator increases when the radius is
decreased so that at a certain point it begins to dominate the other
processes. Usually "in vacuum" this happens at the Planck scale when general
relativity is not a meaningful theory. In the present case, in the
expression of the strength of the gravitational oscillator it is also
present the baryon number $N_{NS}$ of the parent neutron star:%
\begin{equation*}
\rho _{NS}=N_{NS}\frac{m_{Fermions}}{R_{NS}^{3}}
\end{equation*}%
(where $m_{F}$ could be the neutron mass or the quark mass, but this is not
relevant as far as the present paper is concerned). This huge number $N_{NS}$
helps in lowering the critical scale beyond which gravity dominates in a
domain in which classical general relativity and QFT can be trusted.

In the case of a black hole formed during the collapse of a non rotating
neutron star one should solve the Dirac equation in a three dimensional
harmonic potential. However, being the mass of the neutrons as well as the
mass of the quarks much smaller than $\hslash \omega _{G}^{\ast }$ 
\begin{equation*}
m_{quarks}c^{2}\ll \hslash \omega _{G}^{\ast }
\end{equation*}%
the Schrodinger equation can also be used\footnote{%
The corrections due to the Dirac equation are proportional to positive
powers of the ratio $\frac{m_{quarks}c^{2}}{\hslash \omega _{G}^{\ast }}\ll
1 $.}. Therefore, the Fermions live in a three dimensional harmonic
oscillator with frequency $\omega _{G}^{\ast }$ in Eq. (\ref{0QHC}) (an
analysis of \ the "Quantum Hall" behavior of a Fermions gas in a three
dimensional harmonic trap can be found in \cite{HC00}; interesting "harmonic
oscillator" features in black hole physics have been stressed in \cite{NSS06}
\cite{ANSS06} \cite{ACVV01} \cite{ACF06}). Actually, the problem is more
complicated since at densities of the order of the critical density of Eq. (%
\ref{0QHC}) the collapsed matter is well inside its Schwarzschild\ radius.
Therefore, the potential should be an harmonic potential up to the end of
the collapsed matter and a Schwarzschild\ potential outside the collapsed
matter but inside the horizon:%
\begin{eqnarray*}
V_{G}(r) &\approx &I_{1}-\frac{m_{quarks}}{2}\left( \omega _{G}^{\ast
}\right) ^{2}r^{2},r\leq r_{M}, \\
V_{G}(r) &\approx &-G\frac{M_{NS}}{r},\quad r_{M}\leq r\leq r_{G}(M_{NS}), \\
V_{G} &\rightarrow &\infty \quad r>r_{G}(M_{NS}),
\end{eqnarray*}%
where $I_{1}$ is a positive constant, $G$ is the Newton constant, $r_{M}$ is
the radius of the collapsed matter and the Hawking radiation is still
neglected\footnote{%
The effects of Hawking radiation are small at these scale. In any case, they
will be discussed later on.} (so that it can be assumed that the Fermions
are confined to be inside the horizon). This complication does not change
the main new feature of the model: the energy spectrum is still discrete. It
can be also said that the harmonic oscillator part dominates as in all the
successful models of neutron stars the gravitational field outside the
neutron star is not important to determinate the equation of state of the
Fermions living inside the star itself\footnote{%
The inclusion of such deviations is only a technical problem. The
discreteness of the spectrum would not change being, in any case, a
confining potential (see, for instance, \cite{Gia07}). Also the order of
magnitude of the gap should be dominated by the harmonic part which
increases with the decreasing of the radius "strengthening the
incompressibility".}. As it is well known, the eigenvalue and the
degeneracies of a three dimensional harmonic oscillator are%
\begin{equation*}
E_{n}=\hslash \omega _{G}^{\ast }(n+3/2)-I_{1},\quad d(n)=2\left( \frac{%
(n+1)(n+2)}{2}\right)
\end{equation*}%
where the factor of $2$ into the degeneracies is due to the spin degree of
freedom, the negative constant $-I_{1}$ represents a negative contributions
related to the binding energy of the collapsed matter. Let us first suppose
that the Baryon number is such that an integer number of levels is exactly
filled%
\begin{equation}
N_{NS}=\sum_{n}^{n_{\max }}d(n)\approx \left( n_{\max }\right) ^{3}.
\label{0bh0}
\end{equation}%
Because of the energy gap, the Fermions gas becomes incompressible and its
equation of state becomes simply%
\begin{equation*}
\rho =const
\end{equation*}%
which will be used later on to discuss the coupling with the Einstein
equations of such a gas. If the number of baryons does not allow the
complete filling of an integer number of levels, one can write%
\begin{equation*}
N_{NS}=\left( \sum_{n}^{n_{\max }}d(n)\right) +\delta _{n_{\max }},\quad
\delta _{n_{\max }}<d(n_{\max }).
\end{equation*}%
In this case (whose physical features will be discussed in more details in
the next section), the Fermions will form a gas partially compressible.
However, such a gas cannot be compressed beyond the "incompressible core"
constituted by the first $n_{\max }$ fully filled levels. The largest part
of the Fermions is in the incompressible core since%
\begin{equation*}
\frac{\delta _{n_{\max }}}{N_{NS}}<\left( N_{NS}\right) ^{-1/3}\ll 1.
\end{equation*}

\subsection{The general relativistic corrections}

It can be argued that general relativity enhances the arising of "Quantum
Hall phenomenology" of the model: the reason is that general relativity is
more attractive than Newtonian gravity. For a Newtonian star of uniform
density the equilibrium is always possible while in general relativity the
central pressure needed for equilibrium diverges when its gravitational
radius $\left( 2GM\right) /c^{2}$ is greater than $8/9$ of its actual radius
while the pressure far from the origin is quite near to the Newtonian
counterpart. The corrections due to general relativity are "attractive" and
strong at the center of the star. The Fermions perceive a modified harmonic
potential $V_{GR}$ which schematically can be written as follows%
\begin{equation}
V_{GR}=I_{1}-\frac{m_{quarks}}{2}\left( \omega _{G}^{\ast }\right)
^{2}r^{2}-f_{GR}(r)  \label{0pote0}
\end{equation}%
where $f_{GR}$ is a positive function small far from the origin of the star
but which can be large at the origin representing the increased attraction
due to general relativity: for instance%
\begin{equation*}
f_{GR}(r)=\frac{\kappa ^{2}}{r^{\alpha }},\quad \alpha >0,
\end{equation*}%
($\kappa $ and $\alpha $ being real constants) is a reasonable choice to
describe the general relativistic effects on the potential perceived by the
Fermions inside the black hole (the precise form of $f_{GR}(r)$ is not
important, only its qualitative features matter). The effects of such
correction on the wave functions of the Fermions can be evaluated with
perturbation theory. They are very small on the wave functions belonging to
high energy levels since such wave functions are small near the origin where 
$f_{GR}(r)$ is large (actually, such effects can be neglected for all the
wave functions which are not peaked near the origin). The corrections to the
eigenvalues of the Hamiltonian are%
\begin{equation*}
\Delta E_{n}=-\left\langle \psi _{n}\right\vert f_{GR}(r)\left\vert \psi
_{n}\right\rangle
\end{equation*}%
where $\psi _{n}$\ is an eigenfunction belonging to the $n$-th level of the
three dimensional harmonic oscillator. The strength of such corrections
decreases with $n$ 
\begin{equation*}
\partial _{n}\left\vert \Delta E_{n}\right\vert <0
\end{equation*}%
while for small $n$ could be quite strong. The generic result is that the
corrections due to general relativity \textit{enhance the gap} between the
levels (and in particular this fact manifests itself in the eigenfunctions
corresponding to the lowest levels). This means that the corrections due to
general relativity \textit{strengthen the incompressibility} of the Fermions
gas.

It is worth to stress an interesting point. The actual potential perceived
by the particles is the sum of a harmonic oscillator term plus a further
attractive general relativistic correction in which the coupling constants $%
\omega _{G}^{\ast }(t)$ and $\kappa (t)$\ depend adiabatically on time%
\begin{equation}
V_{GR}=I_{1}-\frac{m_{quarks}}{2}\left( \omega _{G}^{\ast }(t)\right)
^{2}r^{2}-\frac{\left( \kappa (t)\right) ^{2}}{r^{\alpha }}.  \label{1pote1}
\end{equation}%
Because of the assumptions made in this paper the Fermions perceive a static
potential. The degeneracies of the highest energy levels are dominated by
the harmonic oscillator part and consequently are constant in time. However,
in the hypothesis of a spherically symmetric collapse, the whole set of
degeneracies is likely not to depend on time. The reason is that any central
potential (besides few integrable exceptions like the harmonic oscillator
itself, the Coulombian potential and so on) has the same degeneracies
related to the spherical symmetry. Therefore, one can roughly divide the
energy levels into the most interior levels which feels the general
relativistic corrections (but whose degeneracies are constant) and the
higher energy levels which are very well approximated by harmonic oscillator
states (so that the corresponding degeneracies are constant as well).

\section{Quantum Hall Effect and black hole entropy}

Many different proposals lead to the same conclusion: the entropy is related
to the area of the horizon. This universality could be related to the
underlying conformal theory living on the boundary which allows to use the
powerful results in \cite{Car86}. It is important to stress also a known but
important fact. The three dimensional BTZ black hole \cite{Banados:1992gq} 
\cite{Banados:1992wn} has entropy as well as Hawking radiation. In such a
case, the degrees of freedom to which the BTZ entropy refers are not related
in an obvious way to gravitational degrees of freedom since in three
dimensions gravity has not local degrees of freedom (this is a highly non
trivial question under active investigation; see \cite{Wi07} and references
therein). To get a reasonable physical picture of the situation (in the
approximation in which gravity is classical), one can imagine to assign the
BTZ entropy to suitable matter degrees of freedom which generate the
singularity at the origin. In the analysis of the spherically symmetric
collapse of a neutron star in four dimensions there are not propagating
gravitational degrees of freedom as well (which would need, at least, a non
trivial quadrupole moment). Therefore, when quantum gravitational effects
are neglected, one can assign the black hole entropy to the degrees of
freedom living inside (and generating) the black hole itself (in the same
way as one actually does in the case of a gas of Fermions living inside a
Newtonian star).

In the case analyzed here, QHE (see \cite{Lau99}, \cite{Wi90}) provides one
with a very natural insight into "why the bulk degrees of freedom are
frozen" so that only boundary excitations are left. The only ingredients are
the presence of a huge number of Fermions (related to the Baryon number
conservation up to energy scale of $10^{-15}$ of the Planck scale) and the
dominance of the classical gravitational attraction.

One possible criticism is that, as the theory of Quantum Hall Effect clearly
stresses \cite{Lau99} \cite{Wi90}, the assumption that the Fermion gas
cannot be compressed anymore depends quite strongly on the "exceptional"
fact that the number of Fermions is such that an integer number of levels is
exactly filled. There is a further argument (which does not depend on the
above "exceptional fact") to see that the present scheme provides with a
qualitative explanation both of why the "dynamical" entropy (this term will
be explained in a moment) is related to the area of the horizon and of how
the bulk degrees of freedom are frozen. However, this argument only works in
the case of Fermions. In the presence of a strong classical gravitational
field the Fermions feel a potential with discrete energy levels. Except the
most interior levels, the others levels and the relative degeneracies are
well approximated by the corresponding quantities of a harmonic oscillator.
The entropy of such a system can be written as follows%
\begin{eqnarray*}
S &=&S_{0}+S_{dyn} \\
S_{0} &=&-\sum_{n}^{n_{\max }-1}p_{n}\log p_{n},\quad S_{dyn}=-p_{n_{\max
}}\log p_{n_{\max }}
\end{eqnarray*}%
where $p_{n}$ is the probability to be in the $n-$th energy level, $n_{\max
} $ is the last partially filled energy level and the reason to split the
entropy into two pieces will be explained in a moment. The present Fermion
gas can be considered as a system at zero temperature since, already for a
neutron star, the Fermi level is much higher than the temperature. For this
reason, the probabilities to be in a given level only depend on the relative
occupation number and on the corresponding degeneracy. It is important in
the present context to split the entropy into two terms because such terms
play different roles. The first term $S_{0}$ is the entropy corresponding to
the interior fully filled levels. Such a term is likely to be constant in
time: due to the gap, even if the whole system could not be in a static
situation, the Fermions inside the fully filled energy level are frozen.
There is no possibility to jump in different energy levels because of the
gap. It is also impossible to jump into different places of the same energy
level because they are fully filled. The first part of the entropy only
depends on the degeneracies of the fully filled levels. In the hypothesis of
the present paper, such degeneracies can be assumed to be constant in time%
\footnote{%
This fact is obvious in the case of a harmonic oscillator since the collapse
simply adiabatically enhances the gap since the time evolution of a
spherical collapse simply enhance the gap keeping the degeneracies. However,
even if the general relativistic corrections are taken into account, the
degeneracies of the energy levels are likely not to depend on time (see the
considerations after Eq. (\ref{1pote1})).}:%
\begin{equation*}
\partial _{t}S_{0}\approx 0.
\end{equation*}%
On the other hand, the last term $S_{dyn}$ corresponds to the last partially
filled energy level. This part is likely not to be constant in time: the
level is only partly filled and the particles living there may interact
jumping into different free places of the same energy level. Because of the
gap, the Fermions inside the partially filled level can only remain in the
same level: the interactions are not able to change the Fermions energies.
Therefore, the only possible excitations should be low energy excitations.
The Fermions living in the last partially filled energy level are not frozen
and $S_{dyn}$ corresponds to the dynamical part of the entropy which can
play an important role during the evolution:%
\begin{equation*}
\partial _{t}S_{dyn}\neq 0.
\end{equation*}%
One can derive a bound for the dynamical entropy: $N_{last}$ (which is the
number of Fermions living in the last partially filled energy level) is
bounded by the degeneracy of the last level:%
\begin{equation}
N_{last}\lesssim d(n_{\max })=(n_{\max }+1)(n_{\max }+2)\approx \left(
N_{NS}\right) ^{2/3}  \label{1area1}
\end{equation}%
where it has been taken into account Eq. (\ref{0bh0}). The total mass of the
gas is proportional to the number of particles ($N_{NS}$ in this case) and
consequently (being the density constant) the volume also is proportional to 
$N_{NS}$. Eq. (\ref{1area1}) tells $N_{last}$ is proportional to the area $%
A_{NS}$ of the horizon of the collapsed neutron star. To get the bound for $%
S_{dyn}$ one can write as usual%
\begin{equation*}
S_{dyn}=\log \Omega
\end{equation*}%
$\Omega $ being the number of possible microscopical configurations
corresponding to the partially filled energy level. A reasonable estimate
for $\Omega $ is the standard binomial expression%
\begin{equation*}
\Omega \approx \frac{\left( d(n_{\max })\right) !}{\left( d(n_{\max
})-N_{last}\right) !\left( N_{last}\right) !}.
\end{equation*}%
Eventually, taking into account the Stirling formula which can be applied in
this case being $N_{NS}$ a very large number, the dynamical part of the
entropy is bounded as follows%
\begin{equation}
\log A_{NS}\lesssim S_{dyn}\lesssim A_{NS}\log A_{NS}  \label{boundentro}
\end{equation}%
which, because of the simplicity of the argument, appears to be a good order
of magnitude estimate strongly suggesting the Bekenstein-Hawking law.

It is also worth to note the above qualitative reasoning works in any
dimension: the reason is that the degeneracies of a $D$-dimensional harmonic
oscillator increases with the energy level label $n$ as%
\begin{equation*}
d_{D}(n)\approx n^{D-1}\Rightarrow \left( n_{D}\right) _{\max }\approx
\left( N_{NS}\right) ^{1/D}\Rightarrow d(\left( n_{D}\right) _{\max
})\approx \left( N_{NS}\right) ^{(D-1)/D}
\end{equation*}%
where $(n_{D})_{\max }$ is the last partially filled energy level in $D$
dimensions and $d_{D}(n)$ is the degeneracy of the $n$-th level in $D$
dimensions. Assuming that the Fermions gas in $D$ dimensions becomes
incompressible one is lead to the conclusion that the dynamical part of the
entropy should be proportional to the area of the collapsed neutron star.

\subsection{Some considerations on Hawking radiation}

Even if for black holes of the mass of a neutron star Hawking radiation is
negligible, it is interesting to note that Hawking radiation should be also
affected by the presence of the discrete spectrum inside the collapsed
matter. Assuming that the black hole evaporation is a real physical process
implies that Hawking particles in some way have to "bring the mass of the
black hole to infinity". Until the final stages, the evaporation is not a
strong gravitational field phenomenon because the black hole mass decreases
slowly with time \cite{Ba81}. The standard Einstein equations with a
suitable matter source can describe how the metric evolves during the
evaporation. As the standard \textit{semiclassical program} has shown (for a
highly incomplete list of references see \cite{Pa80}, \cite{Yo85}, \cite%
{Di92}, \cite{Ka94}, \cite{Pa94}, \cite{Ma95}, \cite{Ma00}, \cite{Ja93}, 
\cite{GKV02} and references therein) the back reaction on the metric due to
a null energy momentum tensor describing the Hawking particles is not able
to stop the evaporation.

It is nevertheless worth to note that it can be shown in the spherical
symmetric four dimensional case \textit{without approximation} that if one
add a trace anomaly term to the energy momentum tensor (allowing, in
principle, arbitrary violations of the energy conditions) the evaporation
process stops \cite{CAVI03}. This conclusion fits quite well with the
results obtained in \cite{Ca01}, \cite{Ah87}.

If the energy spectrum of the particles living inside the horizon is gapped
the Hawking weight could be reduced: as originally found by Hawking \cite%
{Ha74}, the expectation value of the operator number of a Bosonic field of
spin zero, measured by a static observer in the asymptotic future of
Schwarzschild black hole of mass $M$, is 
\begin{equation}
n_{i}(E)=\frac{1}{\exp (E/kT_{H})-1},\quad T_{H}=\frac{\hbar c^{3}}{8\pi kGM}
\label{radhaw}
\end{equation}%
where $k$ is the Boltzmann constant and $E$ is the energy of the particle.
If the spectrum is gapped, the maximum of above expression is obtained for%
\begin{equation*}
n_{i}(E_{gap})=\frac{1}{\exp (E_{gap}/kT_{H})-1}
\end{equation*}%
where $E_{gap}$ is of order in Eq. (\ref{0QHC}). This number, as one can
expect, turns out to be extremely small in the concrete case of black hole
formed during the collapse of a neutron star. The only possibility for the
Hawking particles is to bring outside the black hole the only low energy
degrees of freedom available, namely the gapless boundary excitations.
Therefore, the present scheme suggests that the evaporation process, after
the Hawking particles "have brought away" the gapless deformations of the
boundary, should be highly suppressed by the presence of the gap.

\section{The coupling with Einstein equations}

In \cite{MM04} a very interesting alternative scenario for the final state
of the gravitational collapse has been proposed: the relation of such a
proposal with the present scheme will be discussed in the following. In this
section we will set 
\begin{equation*}
\hbar =1,\quad c=1
\end{equation*}%
while keeping the Newton constant.

One has to describe the interior space time inside the horizon of a black
hole. The metric inside the horizon but outside the collapsed matter is the
Schwarzschild metric. Inside the collapsed matter before classical gravity
begins to dominate the interactions of the standard model, the solution can
be represented by the well known Tolman-Oppenheimer-Volkoff-Snyder solution
(which behaves like a Friedman-Robertson-Walker cosmological solution
matched with the Schwarzschild\ metric). Thus, the interior QHE solution,
whose equation of state is $\rho =const$, has two boundaries: one space-like
boundary separating the standard Tolman-Oppenheimer-Volkoff-Snyder interior
solution from the interior solution in which the equation of state of the
matter changes into an incompressible gas\footnote{%
Schematically, one can write the surface representing the spacelike boundary
can be written as $\tau =\tau ^{\ast }$ where $\tau $ is the proper time of
the Fermions inside the collapsed neutron star and $\tau ^{\ast }$ can be
thought as the time when gravity begins to dominate the other interactions.}%
. The time-like boundary separates the interior QHE solution from the
exterior Schwarzschild metric\footnote{%
Schematically, one can write the surface representing the timelike boundary
can be written as $R=R(\tau )$ where is the radius of the collapsed matter
which, in principle, could depend on time.}. Furthermore, one would like to
find an interior solution which as smoothly as possible matches with the
standard interior Tolman-Oppenheimer-Volkoff-Snyder solution.

One can analyze this problem with the technique of the junction conditions.
Such a technique allows to match two different solutions provided that the
metric is continuous and the discontinuity of the extrinsic curvature is
compensated by a suitable energy momentum tensor $S_{\mu }^{\nu }$ living on
the junction hypersurface. With a strange enough $S_{\mu }^{\nu }$ also
quite different metrics could be matched. One has to search for a junction
in which $S_{\mu }^{\nu }$ is suitable to describe "QHE" features. In the
present context, since $S_{\mu }^{\nu }$ is the candidate to describe the
boundary degrees of freedom of the incompressible gas, it should represent
the classical limit of an energy momentum tensor describing low energy
gapless excitations. This means that one should only allow vanishing or
traceless $S_{\mu }^{\nu }$. This problem can be solved using two results in
the cosmological literature found in a different context (\cite{BGG87} \cite%
{FMM89} \cite{FMM90} \cite{EB01}).

\subsection{The space-like boundary}

Here it will be analyzed how to match the interior "Quantum Hall Metric"
with the standard interior solution describing the collapsed neutron star.
Actually, the would be interior metric describing the quantum hall fluid has
not been found yet. There is indeed some arbitrariness due to the fact that
it is only known that $\rho $ is constant while the pressure is
unconstrained. To overcome this problem one can search for an interior
solution with $\rho =const$ such that the matching with the
Tolman-Oppenheimer-Volkoff-Snyder (henceforth TOVS) solution is as smooth as
possible. Using the results of \cite{FMM89} \cite{FMM90}, one can see that
this can be done \textit{without introducing any} $S_{\mu }^{\nu }$ \textit{%
when the interior metric is the de Sitter one} (in which both $\rho $ and $p$
are constant). It is a very welcome fact that along the space-like boundary
no $S_{\mu }^{\nu }$ is needed\footnote{%
A traceless $S_{\mu }^{\nu }$ is a welcome feature on the timelike boundary
while on the spacelike boundary the interpretation of $S_{\mu }^{\nu }$ as
the energy momentum tensor of gapless excitations would be less clear (since
such excitations would live on a manifold without a time direction).}. This
implies that the present model is quite natural since the matching can be
done in a rather smooth way.

Different interior metrics can indeed be considered since the equation of
state does not determine uniquely the interior solution. On the other hand,
the relations with higher dimensional QHE (discussed in the next sections)
strongly suggest that the interior QHE solution should be described by a
constant curvature metric.

The metric inside the matter during the collapse is a part of the closed
Friedman-Robertson-Walker universe (see, for instance, \cite{Wa84})%
\begin{eqnarray}
ds^{2} &=&a_{F}^{2}(\tau )(-d\tau ^{2}+d\chi ^{2}+\sin ^{2}\chi d\Omega
^{2}),  \label{0FF0} \\
a_{F}(\tau ) &=&a_{0}(1-\cos \tau ),\quad 0\leq \chi \leq \chi _{0}<\frac{%
\pi }{2}.  \notag
\end{eqnarray}%
When $\tau =\pi $ the collapsing star reaches the maximum expansion $%
a_{F}(\pi )=2a_{0}$. The mass $m$ of the collapsing Friedman fluid is
constant during the evolution and reads 
\begin{equation*}
m=\frac{3a_{0}}{2}\left( \chi _{0}-\sin \chi _{0}\cos \chi _{0}\right) .
\end{equation*}%
Due to the gravitational self energy the mass of the external, the mass of
the external Schwarzschild black hole is%
\begin{equation*}
M=a_{0}\sin ^{3}\chi _{0}.
\end{equation*}%
The radius $r(\tau )$ of the collapsing matter evolves as%
\begin{equation}
r(\tau )=a_{F}(\tau )\sin \chi _{0}.  \label{0rad0}
\end{equation}%
The metric in Eq. (\ref{0FF0}) has to be matched with the de Sitter metric.
The suitable coordinates system to write the de Sitter metric is%
\begin{eqnarray*}
ds^{2} &=&a_{dS}^{2}(t)(-dt^{2}+d\chi ^{2}+\sin ^{2}\chi d\Omega ^{2}), \\
a_{dS}(t) &=&\frac{l}{\sin t},
\end{eqnarray*}%
where $l$ is the cosmological length. As it has been already explained, the
matching has to be performed on a space-like hypersurface $\Sigma _{0}$ in
such a way that the energy-momentum tensor $S_{\mu }^{\nu }$ of the
hypersurface vanishes\footnote{%
The possibility to do the matching without any $S_{\mu }^{\nu }$ is a rather
non trivial requirement. The reason is that the equation of state in the
standard phase (in which the pressure can be considered small and the
density increases) is completely different from the equation of state of the
"Quantum Hall phase" (in which the density is constant and the pressure can
be, in principle, arbitrarily large). The fact that de Sitter spacetime
passes this test is a strong confirmation that the approach here proposed is
physical sensible.}.

The matching conditions, as formulated in \cite{Is66}, can be introduced as
follows. Let $\Sigma $ the non null hypersurface on which the matching has
to be performed. Let $\xi ^{\mu }$ the normal to $\Sigma $, and let 
\begin{equation*}
h^{\mu \nu }=g^{\mu \nu }\mp \xi ^{\mu }\xi ^{\nu }
\end{equation*}%
the metric induced on $\Sigma $ (in which the minus sign corresponds to a
space-like $\xi ^{\mu }$ and to time-like $\Sigma $ and the plus sign to the
other case). The matching conditions are that the metric has to be
continuous across $\Sigma $ and 
\begin{eqnarray}
\gamma _{\nu }^{\mu } &=&-8\pi G(S_{\nu }^{\mu }-\frac{1}{2}\delta _{\nu
}^{\mu }trS),  \label{0mmmmm0} \\
\gamma _{\mu \nu } &=&\underset{\varepsilon \rightarrow 0}{\lim }\left[
K_{\mu \nu }(\eta =+\varepsilon )-K_{\mu \nu }(\eta =-\varepsilon )\right]  
\notag
\end{eqnarray}%
where $K_{\mu \nu }$ is the extrinsic curvature of $\Sigma $%
\begin{equation*}
K_{\alpha \beta }=h_{\alpha }^{\mu }h_{\beta }^{\nu }\nabla _{\mu }\xi _{\nu
},
\end{equation*}%
$\eta $\ is the arc length measured along the geodesic orthogonal to $\Sigma 
$. and $S_{\mu }^{\nu }$ is the energy momentum tensor living on $\Sigma $
needed to compensate for the discontinuity of the extrinsic curvature%
\begin{equation*}
S_{\alpha \beta }=h_{\alpha }^{\mu }h_{\beta }^{\nu }S_{\mu \nu }.
\end{equation*}

The jump conditions at the space-like hypersurface $\Sigma _{0}$ (which has
parametric equations $t=t_{0}=const$ and $\tau =\tau _{0}=const$ in the de
Sitter and in Friedman metrics respectively) are (see \cite{FMM90})%
\begin{eqnarray}
a_{F}(\tau _{0}) &=&a_{dS}(t_{0}),\quad S_{\mu }^{\nu }=-\frac{\lambda }{%
4\pi }\delta _{\mu }^{\nu },  \label{0mmm0} \\
\lambda  &=&\left. a_{dS}^{-2}\frac{\partial a_{dS}}{\partial t}\right\vert
_{t=t_{0}}-\left. a_{F}^{-2}\frac{\partial a_{F}}{\partial t}\right\vert
_{\tau =\tau _{0}}.  \label{1M1}
\end{eqnarray}%
Because of the requirements coming from "Quantum Hall Physics", we have to
search for a solution of the equation 
\begin{equation}
\left. a_{dS}^{-2}\frac{\partial a_{dS}}{\partial t}\right\vert
_{t=t_{0}}=\left. a_{F}^{-2}\frac{\partial a_{F}}{\partial t}\right\vert
_{\tau =\tau _{0}}.  \label{2M2}
\end{equation}%
Obviously, such a requirement (which is an input of "Quantum Hall Physics")
was not present in the main references for this section \cite{FMM89} \cite%
{FMM90}. Explicitly, taking into account Eq. (\ref{0mmm0}), Eq. (\ref{2M2})
reads%
\begin{equation*}
\left( 2\frac{a_{0}}{a_{F}(\tau _{0})}-1\right) ^{1/2}=\left( \left( \frac{%
a_{F}(\tau _{0})}{l}\right) ^{2}-1\right) ^{1/2}
\end{equation*}%
so that%
\begin{equation}
a_{F}(\tau _{0})=\left( 2la_{0}\right) ^{1/3}  \label{0mmmm0}
\end{equation}%
which (when inserted in Eq. (\ref{0rad0})) gives a measure of the radius of
the collapsed star when the quantum hall regime sets in.

\subsection{The time-like boundary}

One is left with the problem to match the interior de Sitter solution with
the exterior Schwarzschild solution with a traceless $S_{\mu }^{\nu }$ along
the time-like boundary. A traceless $S_{\mu }^{\nu }$ is a welcome feature
because it would allow the description of the boundary gapless excitations
expected on Quantum Hall grounds. It is convenient, in this case, to use the
following coordinates systems for the interior de Sitter solution $ds_{I}^{2}
$ and for the exterior Schwarzschild solution $ds_{E}^{2}$ respectively%
\begin{eqnarray}
ds_{I}^{2} &=&-(1-k^{2}R^{2})dT^{2}+\frac{dR^{2}}{(1-k^{2}R^{2})}%
+R^{2}d\Omega ^{2},  \label{0DS0} \\
ds_{E}^{2} &=&-(1-\frac{2GM}{R})dT^{2}+\frac{dR^{2}}{(1-\frac{2GM}{R})}%
+R^{2}d\Omega ^{2},  \label{0SCH0} \\
k^{2} &=&\frac{8\pi G}{3}\rho _{0}  \label{0den0}
\end{eqnarray}%
where $\rho _{0}$ is the density of collapsed matter which is of the order
in Eq. (\ref{1QHC}). The results in \cite{BGG87}\footnote{%
in which the authors considered the problem to match the de Sitter solution
to the exterior Schwarzschild solution along a timelike hypersurface.
However, in \cite{BGG87} the authors were not interested in a traceless $%
S_{\mu }^{\nu }$ and considered a different $S_{\mu }^{\nu }$ to describe
the evolution of false vacuum bubbles.} tells that a matching along a
time-like boundary can be achieved. Because of the spherical symmetry, it
can be assumed that the spatial sections of the time-like matching
hypersurface $\Sigma _{(t)}$ are isomorphic to the two sphere so that there
exist a coordinates system in which the induced metric $\left.
ds^{2}\right\vert _{\Sigma _{(\tau )}}$ and $S_{\mu }^{\nu }$ respectively
read:%
\begin{eqnarray}
\left. ds^{2}\right\vert _{\Sigma _{(\tau )}} &=&-d\tau ^{2}+r^{2}(\tau
)d\Omega ^{2}  \notag \\
S^{\mu \nu } &=&\sigma (\tau )\left( U^{\mu }U^{\nu }\right) -\zeta \left(
\tau \right) \left( h^{\mu \nu }+U^{\mu }U^{\nu }\right) ,  \label{0emt0} \\
h^{\mu \nu } &=&g^{\mu \nu }-\xi ^{\mu }\xi ^{\nu }  \notag
\end{eqnarray}%
where, as in the previous subsection, $h^{\mu \nu }$ is the metric induced
on $\Sigma _{(\tau )}$, $\xi ^{\mu }$ is the (space-like) normal to $\Sigma
_{(\tau )}$, $\tau $ is the arc length measured along the time-like geodesic
belonging to $\Sigma _{(\tau )}$, $U^{\mu }$ is the normalized four velocity
of $\Sigma _{(\tau )}$, $\sigma $ is the surface energy density of $\Sigma
_{(\tau )}$, $\zeta \left( \tau \right) $\ is the surface tension and the
conservation of $S_{\mu }^{\nu }$ implies%
\begin{equation*}
\partial _{\tau }\sigma =-2(\sigma -\zeta )\frac{\partial _{\tau }r}{r}
\end{equation*}%
$r(\tau )$ being the proper circumferential radius of the domain wall $%
\Sigma _{(\tau )}$. Unlike the dynamics of false vacuum bubble, here the
requirement to be consistent with a "Quantum Hall picture" tells that $%
S^{\mu \nu }$ has to be chosen traceless (otherwise it would not correspond
to the classical description of boundary gapless degrees of freedom)
therefore one gets%
\begin{eqnarray}
\sigma +2\zeta  &=&0\rightarrow \frac{\partial _{\tau }\sigma }{\sigma }=-3%
\frac{\partial _{\tau }r}{r}\Rightarrow   \label{0co0} \\
\sigma  &=&\frac{\sigma _{0}}{r^{3}},\quad \sigma _{0}>0  \label{0cococo0}
\end{eqnarray}%
where $\sigma _{0}$ is an integration constant which depends on the
microscopic model.

Namely, $\sigma _{0}$ is related to the surface tension (also related to the
energy density of the boundary degrees of freedom) of the incompressible
three dimensional gas (so that it can be assumed to be positive). In the
Quantum Hall case many tools (related to Conformal Field Theory) would come
into play to determine the analogous parameter; in the present case to
determine $\sigma _{0}$ from the microscopic theory appears to be a rather
difficult task. One can therefore deal with $\sigma _{0}$ as a
phenomenological parameter: the following results have a nice interpretation
if compared with the dynamics of false vacuum bubble.

The "equation of motion of the domain wall" that is, the equation which
determines the evolution of $r(\tau )$ can be deduced from the matching
condition (\ref{0mmmmm0}). In particular, the important equation is the
angular $\theta \theta $ component of Eq. (\ref{0mmmmm0})%
\begin{equation}
\gamma _{\theta }^{\theta }=-8\pi GS_{\theta }^{\theta }.  \label{0eqo0}
\end{equation}%
One can see that the $\tau \tau $ component of the matching equations is not
independent on the $\theta \theta $ one provided the conservation of energy
of $S_{\mu }^{\nu }$ is taken into account so that it is enough to deal with
the $\theta \theta $\ component.

It is interesting to note that at a first glance, the right hand side of Eq.
(\ref{0eqo0}) is the same as it appears in the dynamics of false vacuum
bubble (even if in such a case \cite{BGG87} the $S_{\mu }^{\nu }$ \textit{is
not} traceless) 
\begin{equation*}
-8\pi GS_{\theta }^{\theta }=8\pi G\zeta =-4\pi G\sigma ,
\end{equation*}%
where Eqs. (\ref{0emt0}) and (\ref{0co0}) have been taken into account,
because in the dynamics of false vacuum bubble the trace of $S_{\mu }^{\nu }$
cooperates to give, formally, the same result. The important difference
related to the traceless condition will be manifest in a moment.

The standard procedure to compute $\gamma _{\theta }^{\theta }$ is as
follows: because the domain wall is spherically symmetric, on the exterior
Schwarzschild side, the four velocity of any of its point can be written as
follows%
\begin{equation}
U_{(E)}^{\mu }=\left( \frac{\partial _{\tau }r}{1-\frac{2GM}{r}},(1-\frac{2GM%
}{r})\partial _{\tau }t,0,0\right)  \label{1nove1}
\end{equation}%
where it has been taken into account that, on the exterior Schwarzschild
side (\ref{0SCH0}), the coordinate $R$ approaches to $r$ when approaching \
the domain wall. On the interior de Sitter side, the four velocity of any of
the point of the domain wall is%
\begin{equation}
U_{(I)}^{\mu }=\left( \frac{\partial _{\tau }r}{1-k^{2}r^{2}}%
,(1-k^{2}r^{2})\partial _{\tau }t_{(I)},0,0\right)  \label{2nove2}
\end{equation}%
where the notation $t_{(I)}$\ has been introduced to stress that that $%
t_{(I)}$ refers to the interior coordinate. To compute the left hand side of
Eq. (\ref{0eqo0}) the four velocity has to have unit norm in both
coordinates systems, therefore taking into account Eq. (\ref{1nove1}) in the
exterior Schwarzschild coordinates system one has%
\begin{equation*}
(1-\frac{2GM}{r})\partial _{\tau }t=\pm \sqrt{\left( \partial _{\tau
}r\right) ^{2}+1-\frac{2GM}{r}}
\end{equation*}%
while taking into account Eq. (\ref{2nove2}) in the interior de Sitter side
one gets%
\begin{equation*}
(1-k^{2}r^{2})\partial _{\tau }t=\pm \sqrt{\left( \partial _{\tau }r\right)
^{2}+1-k^{2}r^{2}}.
\end{equation*}%
The $\theta \theta $-component of the extrinsic curvature on the exterior
side can be computed 
\begin{equation}
K_{\theta \theta }(ext)=\pm r\sqrt{\left( \partial _{\tau }r\right) ^{2}+1-%
\frac{2GM}{r}}  \label{3nove3}
\end{equation}%
while the $\theta \theta $-component of the extrinsic curvature on the
interior side is%
\begin{equation}
K_{\theta \theta }(int)=\pm r\sqrt{\left( \partial _{\tau }r\right)
^{2}+1-k^{2}r^{2}}.  \label{4nove4}
\end{equation}%
Eventually, Eq. (\ref{0eqo0}) reads%
\begin{equation}
r\left( K_{\theta \theta }(int)-K_{\theta \theta }(ext)\right) =4\pi G\sigma
r^{2}.  \label{00pre00}
\end{equation}%
As it will be shown in a moment, in the dynamical equation for $r(\tau )$
the ambiguity on the relative sign of $K_{\theta \theta }(int)$ and $%
K_{\theta \theta }(ext)$ is not present. This equation can be written as an
ordinary first order equation for $r(\tau )$ bringing on the right hand side 
$K_{\theta \theta }(ext)$ and then squaring (obtaining, at a first glance,
Eq. (5.1) of \cite{BGG87}). However, now an important difference comes into
play. Namely the traceless condition of $S_{\mu }^{\nu }$ together with the
conservation of energy for $S_{\mu }^{\nu }$ itself which imply that $\sigma 
$ is in Eq. (\ref{0cococo0}) while in the description of the dynamics of
false vacuum bubble $\sigma $ can be assumed to be constant. Thus one
obtains the following equation%
\begin{equation}
K_{\theta \theta }(int)-K_{\theta \theta }(ext)=\frac{4\pi G\sigma _{0}}{%
r^{2}}.  \label{0pre0}
\end{equation}%
Equation (\ref{0pre0}) can be written in the standard form%
\begin{eqnarray}
-1 &=&\left( \partial _{\tau }r\right) ^{2}+V(r),  \notag \\
V(r) &=&-\left[ \frac{2GM}{r}+f(r)^{2}\right] ,\quad V(r)\underset{%
r\rightarrow 0}{\rightarrow }-\infty ,\quad V(r)\underset{r\rightarrow
\infty }{\rightarrow }-\infty  \label{potenz00} \\
f(r)^{2} &=&\left[ \left( \frac{M}{4\pi \sigma _{0}}\right) r-\left( \frac{k%
}{8\pi G\sigma _{0}}\right) r^{4}-\frac{2\pi G\sigma _{0}}{r^{2}}\right]
^{2}.  \notag
\end{eqnarray}

It is suggestive to recall that in the dynamics of false vacuum bubble the
effective potential $V_{BV}$, which reads in normalized units%
\begin{equation}
V_{BV}(z)=-\left[ \frac{1-z^{3}}{z^{2}}\right] ^{2}-\frac{\gamma ^{2}}{z},
\label{guth1}
\end{equation}%
has (no matter the choice the parameters) only one maximum so that all the
solutions $r(\tau )$ asymptotically for large $\tau $ approach $+\infty $ or
zero. In particular, neither solutions oscillating between two finite values
of $r$ nor stable constant solutions exist.

\begin{figure}[htb]
\centering
\includegraphics[scale=0.20,angle=-90]{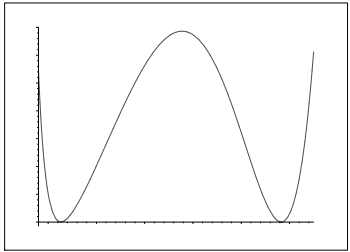} 
\caption{In this figure there is a graph of (minus) the effective potential
for a choice of parameters allowing periodic solutions (a possibility which
is not present in the dynamics of false vacuum bubbles). The graph has been
re-scaled in order to show clearly the "Mexican hat" form of (minus) the
potential.}
\end{figure}
Remarkably enough, in the present case, for suitable choices of the
parameters such solutions exist. In particular, sets of parameters which
provide the potential in Eq. (\ref{potenz00}) with two maxima and a minimum
in between the two maxima exist. It is not possible to write down an
analytic formula for such cases because the potential in Eq. (\ref{potenz00}%
) is involved but a numerical graph unravels this important feature. $V(r)$
can be written as%
\begin{eqnarray}
V(z) &=&-\left( \frac{1}{\sigma _{0}}\right) ^{2}\left[ \frac{\left( \sigma
_{0}\right) ^{2}}{z}+\left[ Bz-Cz^{4}-\frac{\left( \sigma _{0}\right) ^{2}D}{%
z^{2}}\right] ^{2}\right] ,  \label{guth2} \\
B &=&\frac{\left( GM\right) ^{2}}{4\pi G},\quad C=\frac{\left( 2GM\right)
^{4}\rho _{0}}{3},\quad D=\frac{2\pi G}{\left( 2GM\right) ^{2}},
\label{scaling}
\end{eqnarray}%
where Eq. (\ref{0den0}) has been taken into account and the dependence on
the unknown microscopic parameter $\sigma _{0}$ has been displayed (see
Figure 1 for a graph of $-V(z)$ in the range of parameters allowing periodic
solutions). Taking for simplicity $\sigma _{0}=1$ the choice 
\begin{equation*}
B=10,\quad C=\frac{1}{20},\quad D=\frac{1}{80}
\end{equation*}%
does the required job (note that the product of coefficient $B$ and $D$ in
Eq. (\ref{scaling}) is fixed to be $1/8$). One can understand this feature
by comparing $V_{BV}$ in Eq. (\ref{guth1}) with the one in Eq. (\ref{guth2}%
). In the first case the term%
\begin{equation*}
\left[ \frac{1-z^{3}}{z^{2}}\right] ^{2}
\end{equation*}%
has only one zero for $z=1$ while the quadratic term in Eq. (\ref{guth2}) 
\begin{equation*}
\left[ Bz-Cz^{4}-\frac{D}{z^{2}}\right] ^{2}=\frac{1}{z^{4}}\left[
Bz^{3}-Cz^{6}-D\right] ^{2}
\end{equation*}%
can have two zeros\footnote{%
So that (at least when the term $-1/z$ in the potential is not taken into
account) it is clear that $V(z)$ may have two maxima and a minimum in
between.} opening the possibility to have oscillating as well as stable
static solutions living in the local minimum of $-V(z)$: such a possibility
appears to be favored by small values of $\sigma _{0}$. The fact that, once
the Quantum Hall regime sets in, solutions for $r(\tau )$ oscillating around
a local minimum do indeed exist is an interesting new feature of the present
scheme. In Figure 2 there is a schematic Penrose diagram in the interesting
range of parameters. 
\begin{figure}[h]
\centering
\includegraphics[scale=0.37]{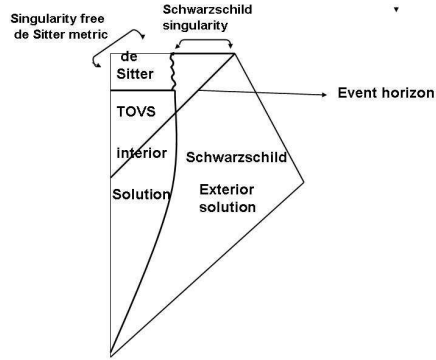}
\caption{Here there is a schematic Penrose diagram in the range of
parameters in which oscillating solutions exist. The line separating the de
Sitter from the Schwarzschild solutions is "wavy" to stress the "Mexican
hat" form of $-V(z)$.}
\end{figure}

\subsection{Connection with the higher dimensional Quantum Hall Effect}

Even if QHE appears as a purely two dimensional phenomenon, its theoretical
structure has been generalized to higher dimensions in \cite{ZJ01} (for a
review, see \cite{KN06}\ and references therein). The basic mathematical
structure needed to achieve such generalizations is a manifold endowed with
a connection acting on spinors and taking value in a Lie algebra. The
curvature of such \textit{Quantum Hall connection} has to be "constant":
namely, the components of the curvature evaluated in a suitable basis of
vielbein have to be constant. The prototype of such manifolds are \textit{%
coset spaces,} namely manifolds diffeomorphic to $G/H$ where $G$ is a Lie
group and $H$ a compact subgroup of dimension $\geq 1$: the spin connection
provides one with a constant background field so that one can choose the
background gauge field to be proportional to the spin connection
generalizing the concept of constant magnetic field\footnote{%
It was found in \cite{ZJ01} that in order to obtain a reasonable
thermodynamic limit with a finite spatial density of particles, one has to
consider very large $SU(\mathbf{2})$ representations. Each particle is then
endowed with an infinite number of $SU(\mathbf{2})$ internal degrees of
freedom. Basically, the reason for this choice is that the authors want to
find a ground state which already has a macroscopic degeneracy (as it
happens in the 2+1 dimensional Quantum Hall effect). In the present case,
this is not an issue: the Fermions live in their own representation (which
is not a large representation of the "internal" Lorentz group of the spin
connection). This is a welcome feature in the present case: an important
ingredient to get the previous entropy bound (\ref{boundentro}) is that the
last partially filled energy level is highly degenerate while the ground
state has not a macroscopic degeneracy.}. Eventually, the Landau problem is
expressed in terms of the covariant derivative of such Quantum Hall
connection and gives rise to a discrete highly degenerate spectrum with a
gap \cite{ZJ01}. It is a highly non trivial self-consistency test of the
present scheme that from the solutions of the Einstein equations fulfilling
the requirements motivated above it naturally emerges the de Sitter metric
which has precisely the characteristic giving rise to higher dimensional
"Quantum Hall" providing one with a spin connection acting on Fermions whose
curvature, as it is well known, is constant. This relation between higher
dimensional Quantum Hall effect and the interior of a black hole is worth to
be further investigated. Furthermore, a connection with non commutative
geometry based on the present scheme and the results in \cite{NSS06} \cite%
{ANSS06} \cite{ACVV01} \cite{ACF06}\ should not be excluded.

\section{Relations with cosmology and gravastars}

It has been shown that the interior solution describing a phase in which the
Fermions gas is incompressible can be chosen to be a de Sitter metric which
is smoothly matched with the standard TOVS solution (describing the interior
of a collapsed neutron star up to energy of the order $10^{-21}$ of the
Planck energy) along the space-like boundary and with the Schwarzschild
solution on the time-like boundary (with a surface energy momentum tensor
describing the boundary gapless excitations expected on Quantum Hall
grounds). In the cosmological literature, interesting models (see, for
instance, \cite{BGG87} \cite{FMM89} \cite{FMM90}) propose that "inside a
black hole a \textit{baby} universe could be generated". The matching with
an interior de Sitter metric is argued to be reasonable on various grounds.
In \cite{BGG87} this scheme represents the evolution of a false vacuum
bubble separated by the true vacuum bubble by a time-like hypersurface. In 
\cite{FMM89} \cite{FMM90} (assuming that the would be quantum theory of
gravity will regularize the divergence of the curvature invariants of
general relativity) the collapsing black hole is matched with a de Sitter
interior inside the horizon on a space-like boundary. In the proposal of 
\cite{BGG87} part of the Schwarzschild singularity is smoothed while in the
proposal of \cite{FMM89} \cite{FMM90} the whole Schwarzschild singularity is
removed. In both cases, instead of the (partially or fully) removed
singularity it is present a "baby universe" in which the inflationary
evolution would arise in a rather natural way. It is therefore interesting
that the conditions to realize such scenario would arise inside a collapsed
neutron star at energy scale of order $10^{-18}$ of the Planck scale.

Recently, an interesting proposal, called \textit{gravastar}, for the final
state of the collapse of a massive star (alternative to the black hole) has
been discussed \cite{MM04} (for a discussion on how it is possible to
distinguish phenomenologically a gravastar from a black hole see, for
instance, \cite{CR07} and references therein). The authors propose as a
final state of the gravitational collapse an incompressible fluid (which
could be a Bose-Einstein condensate) described by an equation of state in
which both $\rho $ and $p$ are constant so that their interior solution is
described by a de Sitter metric. The ambitious idea is that if this would
happen before the formation of the event horizon one could solve the
difficult theoretical problems related with black hole entropy and Hawking
radiation. A point which in the literature on gravastars has not been solved
yet is the precise mechanism giving rise to a Bose-Einstein condensate. On
the other hand, once a neutron star is formed, the theory predicts that for
masses larger than (more or less) five solar masses the neutron star should
collapse to form a black hole: such predictions appear to be quite sound.
Therefore, if the would be mechanism giving rise to gravastar does not come
into play preventing the formation of a neutron star, the formation of a
black hole seems unavoidable. The main idea of the present paper is that the
Fermionic nature of the particles living inside a neutron star together with
the strong gravitational field could give rise to "Quantum Hall
Phenomenology". When the typical order of magnitudes of a neutron star are
taken into account, one recognizes that this would be Quantum Hall Phase
occurs \textit{after the horizon is formed}. The "Quantum Hall
Phenomenology" tells that interior solution is well described by a de Sitter
metric as in \cite{MM04}. However, unlike the results in \cite{MM04}, the
present scheme suggests that "Quantum Hall Phenomenology" together with the
Fermionic nature of the particles living inside the neutron star confirm
that the entropy should be the sum of a frozen constant plus a term
proportional to the area.

\subsection{About the correctness of the assumptions}

The arising of an incompressible gas of Fermions depend on the assumption
that at energy scales up to $10^{-18}$ of the Planck energy quantum
gravitational effects can be neglected and that the standard model can be
fully trusted (so that, for instance, the Baryon number is conserved).
Another key assumption related to the previous one is that at energy scales
lower than $10^{-18}$ of the Planck energy the time scale of the dynamics of
the gravitational field is greater than the typical time scale of the
quantum evolution of the Fermions living inside the neutron star so that the
Fermions reach equilibrium "before the gravitational field changes" (which
is the standard assumption in the theory of the evolution of neutron stars).
If these assumptions are correct, the results of the present paper can be
trusted. About the first assumption, nothing precise can be said since the
final theory of quantum gravity is still lacking. Nevertheless, a comparison
with some analogous situations suggests that such an assumption could be
safe. For instance, if one is studying quantum electrodynamics at an energy
scale which is 18 order of magnitude less than the energy scale at which
quantum effects come into play, classical electrodynamics should be enough
(unless there are very few photons but this is not the present case in which
there should be a huge number of gravitons). Standard dimensional arguments
would suggest that the second assumption also could not be incorrect. The
time derivatives in the Einstein equations appear together with factors of
the Newton constant $G$%
\begin{equation*}
\frac{1}{G}\frac{\partial ^{2}}{\partial t^{2}}
\end{equation*}%
while the time derivatives in Quantum Field Theory acting on the Fermionic
operators appear as follows%
\begin{equation*}
\frac{1}{\hbar }\frac{\partial }{\partial t}
\end{equation*}%
therefore the two evolutions become comparable at the Planck scale. This
standard argument leads to think that (as it happens inside usual collapsing
neutron stars) Fermions reach equilibrium before the gravitational field
changes relevantly so that one can use the Fermions equation of state ($\rho
=const$ in the present case) to solve the Einstein equations. In similar
situations (in which the microscopic time scale is 18 order of magnitude
smaller than the macroscopic time scale) one would say that it is safe to
assume that the microscopic degrees of freedom reach the equilibrium. On the
other hand, quantum gravitational effects could manifest themselves in a
subtle way preventing the Fermions from reaching equilibrium (and making
incorrect the hypothesis made here). If this is the case, it would be a
rather novel type of "low energy" quantum gravitational effect worth to be
further investigated (since, to the best of author's knowledge, no similar
effects of "lacking of equilibrium" in the presence of so different time
scales have been studied).

A possible mechanism preventing the picture here proposed could be to
transform the Fermions into Bosons: in this case, the present picture would
be incorrect. On the other hand, no obvious way to realize that is
available. At energy scale of $10^{-18}$ of the Planck energy the leading
interactions are the strong interactions among the neutrons (or the quarks)
which as it has been already discussed are weaker than the classical
gravitational field inside the black hole.

Various models leading to superfluidity due to the formation of Bosonic
bound states of neutrons via "Bardeen-Cooper-Schrieffer mechanism" inside a
neutron star (see, for a review, \cite{Pet92}) have been proposed. On the
other hand, at the energy scales at which classical gravity dominates the
strong interactions which are responsible for the superfluidity can be
considered as small perturbations. Moreover, the strong interactions are
weaker the higher the energy scale so that the "Bardeen-Cooper-Schrieffer
mechanism"\ could not be effective anymore at density of the order in Eq. (%
\ref{0QHC}). This interesting question is worth to be further investigated.

\section{ Conclusions and perspectives}

It has been argued that the collapse of a black hole formed during the
evolution of a typical neutron star could lead to an incompressible gas of
Fermion well before reaching the Planck scale. The Fermionic nature of the
degrees of freedom together with the strong classical gravitational field
perceived by the Fermions lead to some features typical of Quantum Hall
Effects. The entropy of the gas splits naturally into two terms: a "frozen"
constant (corresponding to the Fermions living in the fully filled discrete
energy levels) and a dynamical term which is bounded by two suitable
functions of the area of the horizon strongly suggesting the
Bekenstein-Hawking area law. The Einstein equations have been solved with
this incompressible fluid and it has been shown that the interior metric
describing the incompressible phase is well described by de Sitter
space-time. The behavior of the matching hypersurface manifests an
interesting dependence on the parameters of the model allowing oscillating
solutions around a local minimum of the effective potential for $r(\tau )$
(the proper circumferential radius of the domain wall $\Sigma _{(\tau )}$),
a peculiar feature of the present model which is related to the gapless
nature of the boundary excitations. The relations with higher dimensional
Quantum Hall effect and with interesting cosmological scenarios have been
pointed: such relations are worth to be further investigated. The case in
which the hypothesis of the present paper do not hold has been shortly
discussed.

\section*{Acknowledgments}

I want to thank J. Zanelli for suggesting to begin this investigation and
for continuous encouraging to complete the work. I thank R. Troncoso for
important suggestions and invaluable support. I want also to thank A.
Giacomini, H. Maeda, J. Oliva and S. Willison for many useful discussions.
The work of F. C. has been partially supported by Proy. FONDECYT N%
${{}^\circ}$%
3070055 and by PRIN SINTESI 2007. The work of F. C. was funded by an
institutional grants to CECS of the Millennium Science Initiative, Chile,
and Fundaci\`{o}n Andes, and also benefits from the generous support to CECS
by Empresas CMPC. 


\begin{thebibliography}{99}
\bibitem{Ba73} J. Bardeen, B. Carter, S. W. Hawking, \textit{Comm. Math. Phys%
}. \textbf{31} (1973) 161.

\bibitem{Be73} J. D. Bekenstein, \textit{Phys. Rev}. \textbf{D} \textbf{7},
2333 (1973).

\bibitem{Ha74} S. W. Hawking, Nature (London) \textbf{243} (1974) 30.

\bibitem{Carl07} S. Carlip, arXiv:0705.3024.

\bibitem{Be81} J. D. Bekenstein,\ \textit{Phys. Rev. }\textbf{D} \textbf{23,}
287 (1981).

\bibitem{tH93} G. 't Hooft, \textquotedblright \textit{Dimensional Reduction
in Quantum Gravity}\textquotedblright\ gr-qc/9310026.

\bibitem{Su95} L. Susskind, \textit{J. Math. Phys} \textbf{36}, 6377 (1995).

\bibitem{Ma97} J. M. Maldacena, \textit{Adv.Theor.Math.Phys}. \textbf{2}
(1998) 231; \textit{Int.J.Theor.Phys}. \textbf{38} (1999) 1113.

\bibitem{Banados:1992gq} M.~Banados, M.~Henneaux, C.~Teitelboim and
J.~Zanelli, 
\textit{Phys.\ Rev.}\ \textbf{D} \textbf{48} (1993) 1506.

\bibitem{Banados:1992wn} M.~Banados, C.~Teitelboim and J.~Zanelli, 
\textit{Phys.\ Rev.\ Lett}.\ \textbf{69}, 1849 (1992).

\bibitem{Myu98} Y.S. Myung, \textit{Phys.Rev.} \textbf{D59} (1999) 044028.

\bibitem{Lau99} R. B. Laughlin, \textit{Rev. Mod. Phys.} \textbf{71} (1999),
863.

\bibitem{RW05} S. P. Robinson, F. Wilczek, \textit{Phys.Rev.Lett}. \textbf{96%
} (2006) 231601.

\bibitem{HC00} Tin-Lin Ho, C. V. Ciobanu, \textit{Phys. Rev. Lett.} \textbf{%
85}, 4648 (2000).

\bibitem{NSS06} P. Nicolini, A. Smailagic, E. Spallucci, \textit{Phys.Lett}. 
\textbf{B632} (2006) 547.

\bibitem{ANSS06} S. Ansoldi, P. Nicolini, A. Smailagic, E. Spallucci, 
\textit{Phys.Lett.} \textbf{B645} (2007) 261.

\bibitem{ACVV01} G. L. Alberghi, R. Casadio, G. P. Vacca, G. Venturi, 
\textit{Phys.Rev}. \textbf{D64} (2001) 104012.

\bibitem{ACF06} G.L. Alberghi, R. Casadio, D. Fazi, \textit{Class.Quant.Grav}%
. \textbf{23} (2006) 1493.

\bibitem{Gia07} R. Giachetti, E. Sorace, arXiv:0706.0127.

\bibitem{Car86} J. A. Cardy, \textit{Nucl. Phys.} \textbf{B270} (1986) 186.

\bibitem{Wi07} E. Witten, "Three-Dimensional Gravity Revisited"
arXiv:0706.3359.

\bibitem{Wi90} F. Wilczek, \textit{Fractional Statistics and Anyon
Superconductivity}, World Scientific (1990).

\bibitem{Ba81} J. Bardeen, Phys. Rev. Lett. \textbf{46} (1981) 382.

\bibitem{Pa80} D. N. Page, \textit{Phys. Rev. Lett}. \textbf{44} (1980) 301; 
\textit{Phys. Rev}. \textbf{D25} (1982) 1499.

\bibitem{Yo85} J. W. York, \textit{Phys. Rev}. \textbf{D} \textbf{31} (1985)
775.

\bibitem{Di92} R. Dijkgraaf, H. Verlinde, E. Verlinde, \textit{Nucl. Phys}. 
\textbf{B} \textbf{371} (1992) 269.

\bibitem{Ka94} D. I. Kazakov, S. N. Solodukhin \textit{Nucl. Phys.} \textbf{B%
} \textbf{429} (1994) 153.

\bibitem{Pa94} R. Parentani, T. Piran, Phys. Rev. Lett. \textbf{73}
(1994)2805.

\bibitem{Ma95} S. Massar, \textit{Phys. Rev.} \textbf{D} \textbf{52} (1995)
5861.

\bibitem{Ma00} S. Massar, R. Parentani, \textit{Nucl. Phys.} \textbf{B} 
\textbf{575} (2000) 333.

\bibitem{Ja93} I. Jack, D. R. T. Jones, J. Panvel, \textit{Nucl. Phys.} 
\textbf{B} \textbf{393} (1993) 95.

\bibitem{GKV02} D. Grumiller, W. Kummer, D. V. Vassilevich, Phys. Rept. 
\textbf{369} (2002) 327 and references therein.

\bibitem{CAVI03} F. Canfora, G. Vilasi, \textit{JHEP} \textbf{0312} (2003)
055.

\bibitem{Ca01} R. Casadio, \textit{Phys.Lett.} \textbf{B} \textbf{511}
(2001) 285.

\bibitem{Ah87} Y. Aharonov, A. Casher, S. Nussinov, \textit{Phys. Lett.} 
\textbf{B} \textbf{191} (1987) 51.

\bibitem{MM04} P. O. Mazur, E. Mottola, \textit{Proc. Nat. Acad. Sci.} 
\textbf{101} (2004) 9545.

\bibitem{BGG87} S. K. Blau, E. I. Guendelman, A. H. Guth, \textit{Phys. Rev.}
\textbf{D55} (1987), 1747.

\bibitem{FMM89} V. P. Frolov, M. A. Markov, V. F. Mukhanov, \textit{Phys.
Lett.} \textbf{B216} (1989), 272.

\bibitem{FMM90} V. P. Frolov, M. A. Markov, V. F. Mukhanov, \textit{Phys.
Rev.} \textbf{D41} (1990), 383.

\bibitem{EB01} D. A. Easson, R. H. Brandenberger, \textit{JHEP} \textbf{0106}
(2001) 024.

\bibitem{Wa84} R. M. Wald, \textit{General Relativity}, Chicago University
press (1984).

\bibitem{Is66} W. Israel, \textit{Nuovo Cimento} \textbf{44B}, 1 (1966); 
\textbf{48B}, 463(E) (1967).

\bibitem{ZJ01} S. C. Zhang, J. Hu, \textit{Science} \textbf{294} (2001) 823.

\bibitem{KN06} D. Karabali, V.P. Nair, \textit{J.Phys}. \textbf{A39} (2006)
12735.

\bibitem{CR07} C. B. M. H. Chirenti, L. Rezzolla, \textit{How to tell a
gravastar from a black hole}, gr-qc/0706.1513.

\bibitem{Pet92} C. J. Pethick, \textit{Rev. Mod. Phys}. \textbf{64} (1992),
1133.
\end{thebibliography}
\end{document}